\documentclass[conference,9pt]{IEEEtran}

\usepackage{cite}
\usepackage{amsmath,amssymb,amsfonts}
\usepackage[ruled]{algorithm2e}
\usepackage{graphicx}
\usepackage{textcomp}
\usepackage{booktabs}
\usepackage{xcolor}
\usepackage{hyperref}
\def\BibTeX{{\rm B\kern-.05em{\sc i\kern-.025em b}\kern-.08em
    T\kern-.1667em\lower.7ex\hbox{E}\kern-.125emX}}

\newcommand{\ci}[1]{\tiny{$\pm${#1}}}

\newcommand{\R}{\mathbb R}

\newcommand{\nN}{\mathcal N}
\newcommand{\dd}{\mathrm d}
\newcommand{\what}{\widehat}

\newcommand{\dpkl}[2]{\left (#1\middle\| #2\right)}
\newcommand{\dkl}[2]{D_{\text{KL}}\dpkl{#1}{#2}}

\newcommand{\n}[1]{\|#1\|}

\DeclareMathOperator*{\argmin}{arg\,min}
\newcommand{\p}[1]{\left({#1}\right)}

\newcommand{\bs}[1]{\left [{#1}\right]}

\begin{document}

\title{E1 TTS: Simple and Fast Non-Autoregressive TTS}

\author{
    \IEEEauthorblockN{Zhijun Liu$^{1}$ \quad Shuai Wang$^{2, 1}$ \quad Pengcheng Zhu\textsuperscript{3} \quad 
    Mengxiao Bi\textsuperscript{3} \quad 
    Haizhou Li$^{1,2}$}
    \IEEEauthorblockA{
        \textit{
            $^1$School of Data Science, $^2$Shenzhen Research Institute of Big Data
        }\\
        \textit{
            The Chinese University of Hong Kong, Shenzhen, Guangdong, P.R. China
        }
        \\\textsuperscript{3}\textit{Fuxi AI Lab, NetEase Inc., Hangzhou, China} \\
        zhijunliu1@link.cuhk.edu.cn \quad wangshuai@cuhk.edu.cn
    }
}
\maketitle

\begin{abstract}
This paper introduces Easy One-Step Text-to-Speech (E1 TTS), an efficient non-autoregressive zero-shot text-to-speech system based on denoising diffusion pretraining and distribution matching distillation. The training of E1 TTS is straightforward; it does not require explicit monotonic alignment between the text and audio pairs. The inference of E1 TTS is efficient, requiring only one neural network evaluation for each utterance. Despite its sampling efficiency, E1 TTS achieves naturalness and speaker similarity comparable to various strong baseline models. Audio samples are available at \href{https://e1tts.github.io/}{e1tts.github.io}.
\end{abstract}


\section{Introduction}

Non-autoregressive (NAR) text-to-speech (TTS) models \cite{TTSSurvey} generate speech from text in parallel, synthesizing all speech units simultaneously. This enables faster inference compared to autoregressive (AR) models, which generate speech one unit at a time. Most NAR TTS models incorporate duration predictors in their architecture and rely on alignment supervision~\cite{DeepVoice,FastPitch,FastSpeech}. Monotonic alignments between input text and corresponding speech provide information about the number of speech units associated with each text unit, guiding the model during training. During inference, learned duration predictors estimate speech timing for each text unit.

Several pioneering studies~\cite{VARA-TTS,Flow-TTS} have proposed implicit-duration non-autoregressive (ID-NAR) TTS models that eliminate the need for alignment supervision or explicit duration prediction. These models learn to align text and speech units in an end-to-end fashion using attention mechanisms, implicitly generating text-to-speech alignment.

Recently, several diffusion-based~\cite{ScoreSDE} ID-NAR TTS models~\cite{E3TTS,SimpleTTS2,SimpleSpeech,E2TTS,Seed-TTS,DiTToTTS,Mapache} have been proposed, demonstrating state-of-the-art naturalness and speaker similarity in zero-shot text-to-speech~\cite{YourTTS}. However, these models still require an iterative sampling procedure taking dozens of network evaluations to reach high synthesis quality. Diffusion distillation techniques~\cite{BlogDistillation} can be employed to reduce the number of network evaluations in sampling from diffusion models. Most distillation techniques are based on approximating the ODE sampling trajectories of the teacher model. For example, ProDiff~\cite{ProDiff} applied Progressive Distillation~\cite{ProgressiveDistillation}, CoMoSpeech~\cite{CoMoSpeech} and FlashSpeech~\cite{FlashSpeech} applied Consistency Distillation~\cite{ConsistencyModel}, and VoiceFlow~\cite{VoiceFlow} and ReFlow-TTS~\cite{ReFlowTTS} applied Rectified Flow~\cite{RectifiedFlow}. Recently, a different family of distillation methods was discovered~\cite{DiffInstruct,DMD2}, which directly approximates and minimizes various divergences between the generator's sample distribution and the data distribution. Compared to ODE trajectory-based methods, the student model can match or even outperform the diffusion teacher model~\cite{DMD2}, as the distilled one-step generator does not suffer from error accumulation in diffusion sampling.

In this work, we distill a diffusion-based ID-NAR TTS model into a one-step generator with recently proposed distribution matching distillation~\cite{DiffInstruct,DMD2} method. The distilled model demonstrates better robustness after distillation, and it achieves comparable performance to several strong AR and NAR baseline systems.

\begin{figure}[!ht]
    \vspace{-20pt}
    \centerline{\includegraphics[scale=1.2]{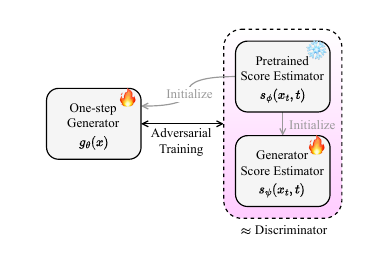}}
    \vspace{-20pt}
    \caption{Distribution matching distillation (DMD) of diffusion models is summarized in this overview. The pretrained score estimator serves to initialize both the one-step generator and the score estimator for the generated samples. Following initialization, the generator is optimized using DMD in a manner analogous to adversarial training.}
    \label{fig:dmd}
\end{figure}

\begin{figure*}[!t]
    \centerline{\includegraphics{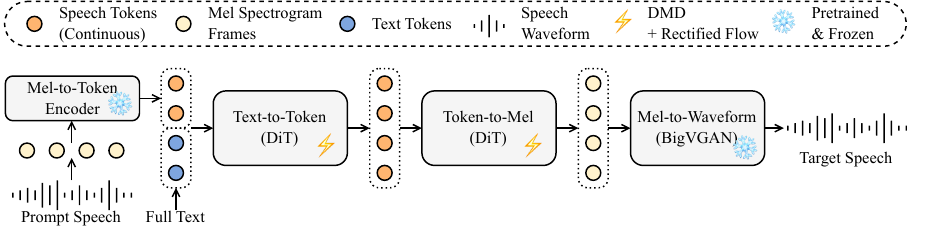}}
    \caption{
        An overview of the E1 TTS inference pipeline in prompted text-to-speech:
        (1)~The reference speech Mel spectrogram is encoded into speech tokens.
        (2)~A Diffusion Transformer (DiT) generates all speech tokens given the prompt speech tokens and the prompt and target text.
        (3)~Another DiT model generates the Mel spectrogram given the generated speech tokens.
        (4)~A neural vocoder converts the input Mel spectrogram to the target waveform.
    }
    \label{fig:e1tts}
\end{figure*}

\section{Background}

\subsection{Distribution Matching Distillation}

Consider a data distribution $p(x)$ on $\R^d$. We can convolve the density $p(x)$ with a Gaussian perturbation kernel $q_t(x_t | x) = \mathcal{N}(x_t; \alpha_t x, \sigma_t^2 \mathbf{I}_d)$ to obtain the perturbed density $p_t(x_t) := \int p(x) q_t(x_t | x) \dd x$, where $\alpha_t, \sigma_t > 0$ define the signal-to-noise ratio at each time $t \in [0, 1]$. Various formulations of diffusion models exist in the literature~\cite{RectifiedFlow,ScoreSDE}, most of which are equivalent to learning a neural network that approximates the score function $s_p(x_t, t) :=\nabla_{x_t} \log p_{t}(x_t)$ at each time $t$.

Now, consider a generator function $g_\theta(z): \R^d \to \R^d$ that takes in random noise $Z \sim \mathcal{N}(0, \mathbf{I}_d)$ and outputs fake samples $\what{X} := g_\theta(Z)$ with distribution $q_\theta(x)$. Several studies~\cite{VSD,DiffInstruct} have discovered that if we can obtain the two score functions $s_p(x) := \nabla_x \log p(x)$ and $s_{q} (x) := \nabla_{x} \log q_\theta(x)$, we can compute the gradient of the following KL divergence:
\begin{equation}
\nabla_\theta \dkl{q_\theta(x)}{p(x)} = E\bs{\p{s_{q}(\what X) - s_p(\what X)}\frac{\partial g_\theta(Z)}{\partial \theta}}.
\end{equation}
However, obtaining $s_p(x)$ and $s_{q}(x)$ directly is challenging. Instead, we can train diffusion models to estimate the score functions $s_p(x_t, t)$ and $s_{q}(x_t, t)$ of the perturbed distributions $p_t(x_t)$ and $q_{\theta, t}(x_t) := \int q_\theta(x) q_t(x_t | x) \dd x$. Consider the following weighted average of KL divergence at all noise scales~\cite{DiffInstruct,VSD}:
\begin{equation}
D_\theta := E_{t \sim p(t)} \bs{w_t \dkl{q_{\theta, t}(x_t)}{p_t(x_t)}},
\end{equation}
where $w_t \ge 0$ is a time-dependent weighting factor, and $p(t)$ is the distribution of time. Let $W \sim \mathcal{N}(0, \mathbf{I}_d)$ be an independent Gaussian noise, and define $\what{X}_t := \alpha_t \what X + \sigma_t W$. Then, the gradient of the weighted KL divergence can be computed as:
\begin{equation}
\nabla_\theta D_\theta  = E_{t \sim p(t)}\bs{w_t \alpha_t\p{s_{q}(\what X_t, t) - s_p(\what X_t, t)}\frac{\partial g_\theta(Z)}{\partial \theta}}.
\label{equ:IKL}
\end{equation}

Given a pretrained score estimator $s_\phi(x_t, t) \approx s_p(x_t, t)$, the procedure to distill it into a single-step generator $g_\theta$ is described in Algorithm \ref{alg:DMD}.
\begin{algorithm}
\SetAlgoLined

\textbf{Input:} Pretrained score estimator $s_\phi(x_t, t)$ that approximates score $s_p(x_t, t)$ of perturbed real data density $p_t(x_t)$.

\textbf{Output:} Single-step generator $g_\theta$ with sample distribution $q_\theta(x) \approx p(x)$.

Initialize the one-step generator $g_\theta: \R^d \to \R^d$.

Initialize score estimator $s_\psi(x_t, t)$ by $\psi \leftarrow \phi$.

\Repeat{convergence}{
    1. Approximate $\nabla_\theta D_\theta$ by replacing $(s_q, s_p)$ with their neural network estimator $(s_\psi, s_\phi)$ in Equation \ref{equ:IKL}. And update $\theta$ to minimize $D_\theta$ with gradient $\nabla_\theta D_\theta$.
    
    2. Draw samples from $g_\theta$ and optimize $s_\psi$ with the denoising score matching loss on the generated samples.
}

\caption{Distillation of pretrained score estimator into single-step generator~\cite{DiffInstruct,DMD2}}
\label{alg:DMD}
\end{algorithm}

Although the generator $g_\theta$ can be randomly initialized in theory, initializing $g_\theta$ with $s_\phi$ leads to faster convergence and better performance\cite{DiffInstruct}. Several studies~\cite{LCMLoRA,InnateOneStep} have discovered that pretrained diffusion models already possess latent one-step generation capabilities. Moreover, it is possible to convert them into one-step generators by tuning only a fraction of the parameters~\cite{LCMLoRA,InnateOneStep}, such as the normalization layers.

While distribution matching distillation resembles generative adversarial networks (GANs)~\cite{GAN} in its requirement for alternating optimization, it has been empirically observed~\cite{DMD2} to be significantly more stable, requiring minimal tuning and avoiding the mode collapse issue that often hinders GAN training.

\subsection{Rectified Flow}

Rectified Flow~\cite{RectifiedFlow} is capable of constructing a neural ordinary differential equation (ODE):
\begin{equation}
    \dd Y_t = v(Y_t, t) \dd t, \quad t \in [0, 1],
\end{equation}
that maps between two random distributions $X_0 \sim \pi_0$ and $X_1 \sim \pi_1$, by solving the following optimization problem:
\begin{equation}
    \label{equ:reflow}
    v(x_t, t) := \argmin_{v} E\n{v\p{\alpha_t X_1 + \sigma_t X_0, t} - (X_0 - X_1)}_2^2,
\end{equation}
where $\alpha_t = t$ and $\sigma_t = (1 - t)$. In the special case where $X_0 \sim \nN(0, \mathbf I_d)$ and $X_0 \perp X_1$, the drift $v(x_t, t)$ is a linear combination of the score function $s(x_t, t) = \nabla_{x_t} \log p_t (x_t)$ and $x_t$, where $X_t := \alpha_t X_1 + \sigma_t X_0$:
\begin{equation}
    \label{equ:svequ}
    s(x_t, t) = - \frac{1 - t}{t} v(x_t, t) - \frac{1}{t}x_t.
\end{equation}
In the experiments, we trained all our diffusion models with the Rectified Flow loss in Equation \ref{equ:reflow}. Equation \ref{equ:svequ} allows us to apply DMD to Rectified Flow models.
\section{E1 TTS}

E1 TTS is a cascaded conditional generative model, taking the full text and partially masked speech as input, and outputs completed speech. The overall architecture is illustrated in Figure~\ref{fig:e1tts}. E1 TTS is similar to the acoustic model introduced in~\cite{ARDiT} with the modification that all speech tokens are generated simultaneously in the first stage. Further more, we applied DMD to convert the two diffusion transformers~(DiTs)~\cite{DiT} to one-step generators, removing all iterative sampling from the inference pipeline. We will describe the components in the system in the following sections.

\subsection{The Mel Spectrogram Autoencoder}

Directly training generative models on low-level speech representations such as Mel spectrograms~\cite{E2TTS} and raw waveforms~\cite{E3TTS} is resource-consuming due to the long sequence lengths. We build a Mel spectrogram autoencoder with a Transformer encoder and a Diffusion Transformer decoder. The encoder takes log Mel spectrograms and outputs continuous tokens in $\R^{32}$ at a rate of approximately 24Hz. The decoder is a Rectified Flow model that takes speech tokens as input and outputs Mel spectrograms. The encoder and decoder are jointly trained with a diffusion loss and a KL loss to balance rate and distortion. The Mel spectrogram autoencoder is fine-tuned for the case where part of the spectrogram is known during synthesis to enhance its performance in speech inpainting. For the decoder, we appended layers of 2D convolutions after the transformer blocks to improve its performance on spectrograms. Please refer to~\cite{ARDiT} for further details regarding the training process and model architecture.

\subsection{Text-to-Token Diffusion Transformer}

\begin{figure}[!ht]
    \vspace{-10pt}
    \centerline{\includegraphics[scale=1.2]{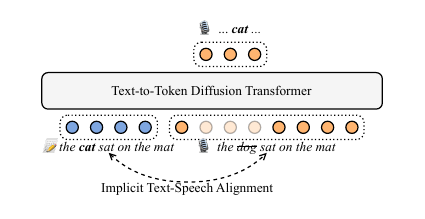}}
    \vspace{-15pt}
    \caption{
        Illustration of the Text-to-Token Diffusion Transformer performing text-based speech editing. The model takes concatenated text and noised speech tokens as input, and predicts the masked speech tokens for the replaced text by predicting the score function. The model implicitly aligns text and speech modalities without token-to-token alignment information.
    }
    \label{fig:maskgen}
\end{figure}

The Text-to-Token DiT is trained to estimate the masked part of input speech tokens given the full text. During training, the sequence of speech tokens is randomly split into three parts: the prefix part, the masked middle part, and the suffix part. We first sample the length of the middle part uniformly, and then we sample the beginning position of the middle part uniformly. With 10\% probability we mask the entire speech token sequence.

\begin{figure}[!ht]
    \centerline{\includegraphics[scale=1.32]{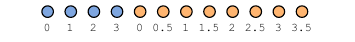}}
    \vspace{-10pt}
    \caption{Token position indices in the Text-to-Token DiT.}
    \vspace{-10pt}
    \label{fig:posemb}
\end{figure}

We adopted rotary positional embedding (RoPE)~\cite{Roformer} in all Transformer blocks in E1 TTS. For the Text-to-Token model, we designed the positional embedding to promote diagonal alignment between text and speech tokens, as illustrated in Figure~\ref{fig:posemb}. With RoPE, each token is associated with a position index, and the embeddings corresponding to the tokens are rotated by an angle proportional to their position index. For text tokens, we assign them increasing integer position indices. For speech tokens, we assign them fractional position indices, with an increment of $n_{\text{text}} / n_{\text{speech}}$. This design results in an initial attention pattern in the form of a diagonal line between text and speech. Similar designs have proven effective in other ID-NAR TTS models~\cite{ParaNet, VARA-TTS}.

\subsection{Duration Modeling}

Similar to most ID-NAR TTS models, E1 TTS requires the total duration of the speech to be provided during inference. We trained a duration predictor similar to the one in~\cite{VoiceBox}. The rough alignment between text and speech tokens is first obtained by training an aligner based on RAD-TTS~\cite{RAD-TTS}. Then a regression-based duration model is trained to estimate partially masked durations. The duration model takes the full text (phoneme sequence in our case) and partially observed durations as input, then predicts unknown durations based on the context. We observed that minimizing the L1 difference in total duration~\cite{Flow-TTS,VARA-TTS} works better than directly minimizing phoneme-level durations, resulting in a lower total duration error.

\subsection{Inference}

The inference process for text-based speech editing~\cite{DiffVoice} with E1~TTS involves several steps. First, the original text and speech are force-aligned to obtain the original phoneme durations. Next, the duration predictor is fed the target phoneme sequence and the original durations of unedited phonemes to estimate the total duration of the target speech. The original speech is then encoded into speech tokens, and the old tokens corresponding to the edited part are removed. New noise tokens are inserted to match the estimated target duration. Finally, the target text and partially masked target speech tokens are fed to the Text-to-Token DiT to obtain the reconstructed speech tokens, resulting in the edited speech output that matches the target text while preserving the original speech characteristics in the unedited parts. The procedure for zero-shot text-to-speech synthesis with E1 TTS works similarly.

For simpler tasks such as single-speaker TTS with only text input, we can remove the force-alignment step and the duration predictor from the inference pipeline. In this case, the total duration of the synthesized speech can be estimated by assuming it is a fixed multiple of the input text length as done in~\cite{ParaNet}. Or we can train a total-duration predictor following~\cite{Flow-TTS,VARA-TTS}, which does not require text unit durations as supervision.

\section{Experiments and Results}


\subsection{Setup}

\textbf{Datasets:} All components of the evaluated E1 TTS model were trained on the LibriTTS~\cite{LibriTTS} dataset, which is a multi-speaker English speech corpus containing 585 hours of speech audio from over 2,300 speakers. We used an open-source BigVGAN~\cite{BigVGAN} checkpoint\footnote{\texttt{github.com/NVIDIA/BigVGAN}  \texttt{(bigvgan\_24khz\_100band)}} to generate 24 kHz speech waveforms from Mel spectrograms.

\textbf{Baselines:} We used the open-source StyleTTS~2\footnote{\texttt{github.com/yl4579/StyleTTS2 (StyleTTS2-LibriTTS)}} and CosyVoice\footnote{\texttt{github.com/FunAudioLLM/CosyVoice (CosyVoice-300M)}} as NAR and AR TTS baseline systems. We utilized their officially provided model weights and inference code for evaluation. Another baseline system is ARDiT-TTS$_\text{B=1, DMD}$ from~\cite{ARDiT}, which has a similar design and shares the Mel-to-Token encoder and Token-to-Mel decoder with E1 TTS. We also compare the distilled model E1~TTS$_\text{DMD}$ with E1~TTS$_\text{ODE}$. E1~TTS$_\text{ODE}$ directly samples the teacher diffusion model through ODE samplers. In ODE sampling, we take 128 Euler steps for the Text-to-Token DiT, and we take 32 steps for the Token-to-Mel DiT.

\textbf{Metrics:} For objective evaluations, we report the Speaker Encoding Cosine Similarity (SECS) and the Word Error Rate (WER) of the generated samples. The {Whisper-medium} model is employed for evaluating WER, while the {WavLM-large} model, fine-tuned on the speaker verification task, is used for evaluating SECS. For subjective evaluations, we conducted MUSHRA tests without hidden reference and anchors to assess speech naturalness~(NAT) and speaker similarity~(SIM). For all evaluations, the generated audios are downsampled to 16kHz. Our evaluation test set and source code can be found in the online supplement.

\subsection{Training Details}

We trained all models using the AdamW optimizer with $\beta_1 = 0.9$ and $\beta_2 = 0.95$, with a constant learning rate of 0.0001. For evaluation, we always use exponential moving averaged weights with a decay rate of 0.9999. The EMA rate has a significant impact on sample quality~\cite{EDM2}. For the DMD training of the DiT models, we adopted the two-timescale update rule~(TTUR)\cite{TTUR}. The generators are updated once per 10 updates of the score model. We observed that TTUR stabilized DMD training significantly as discovered in \cite{DMD2}.

The Mel-to-Token Encoder was trained jointly with the Token-to-Mel DiT for 600k steps. Then the encoder was frozen, and we further trained the Text-to-Mel DiT decoder with DMD for 20k generator steps. The Text-to-Token DiT was trained for 800k steps. Then we further trained it with DMD for 80k generator steps. In all training stages, the batch sizes are dynamic and there are approximately 10 minutes of speech audio in each batch.

\subsection{Zero-Shot Text-to-Speech}

\label{sec:zstts}

Following the evaluation protocol in \cite{ARDiT,UniCATS}, we evaluate the zero-shot TTS performance of all models on {test set B}, derived from the {test-clean} subset of LibriTTS, containing 500 test cases. The test set includes 37 speech prompts from different speakers, each with a duration of approximately 3 seconds. We report the average WER and SECS of all 500 test cases. Note that the SECS scores here are computed between the models' outputs and the speech prompts. For the subjective evaluations, we collected scores from 20 participants well-versed in English, with each participant randomly rating 10 out of 200 randomly selected test cases. The results can be found in Table~\ref{table:zstts}.

\begin{table}[!h]
\caption{Results of zero-shot text-to-speech.}
    \label{table:zstts}
    \centering
    \begin{tabular}{lcccc}
        \toprule
        Model             & NAT$\uparrow$ & SIM$\uparrow$ & WER(\%)$\downarrow$ & SECS$\uparrow$ \\
        \midrule
        Ground Truth & 88.5\ci{2.1} & 75.2\ci{4.1} & 2.02 & 0.675 \\
        Mel Reconstruct & --- & --- & 1.89 & 0.663 \\
        Token Reconstruct$_\text{ODE}$ & --- & ---  & 2.03 & 0.638\\
        Token Reconstruct$_\text{DMD}$ & --- & ---  & 2.02 & 0.640\\
        \midrule 
        StyleTTS 2~\cite{StyleTTS2} & 75.8\ci{2.6} & 60.7\ci{3.8} & 1.76 & 0.410\\
        CosyVoice~\cite{CosyVoice} & 74.4\ci{2.8} & 73.1\ci{2.5} & 2.19 & 0.648\\
        ARDiT$_{\text{B=1, DMD}}$~\cite{ARDiT} & 81.0\ci{2.4} & 85.6\ci{2.0} & 1.88 & 0.614\\
        E1 TTS$_{\text{ODE}}$ & 60.0\ci{4.2} & 68.6\ci{4.3} & 3.28 & 0.608\\
        E1 TTS$_{\text{DMD}}$ & 81.1\ci{2.3} & 84.7\ci{2.2} & 1.92 & 0.620\\
        \bottomrule
        \end{tabular}
\end{table}

\subsection{Text-based Speech Inpainting}

Text-based speech inpainting regenerates a masked speech segment while keeping the corresponding text unchanged. We evaluate the performance of speech inpainting on the same test set described in Section \ref{sec:zstts}. In this experiment, the models are tasked with generating the middle one-third of all 500 utterances, given the full texts and ground truth total durations. Note that WER and SECS were evaluated using the entire audio samples. The SECS scores here are computed between the models' outputs and the original speech. Additionally, we conducted a two-alternative forced choice~(2AFC) test with 10 listeners each rating 50 test cases. We report the preference rate for the model-generated speech over the vocoder-reconstructed speech. E1~TTS$_\text{DMD}$ can achieve comparable performance as the strong baseline model ARDiT$_\text{B=1,DMD}$.

\begin{table}[!h]
    \centering
    \caption{Results on the speech inpainting task.}
    \label{table:speech_editing}
    \begin{tabular}{lccc}
        \toprule
        Model &{Preference(\%)$\uparrow$} & {WER(\%)$\downarrow$} & {SECS$\uparrow$}\\
        \midrule
        Mel Reconstruct & --- & 1.89 & 0.971\\
        Token Reconstruct$_\text{DMD}$ & --- & 2.02 & 0.930\\
		\midrule
		ARDiT$_{\text{B=1, DMD}}$ & 46.2 & 1.89 & 0.789\\
        E1 TTS$_{\text{DMD}}$ & 46.7 & 2.00 & 0.801\\
        \bottomrule
	\end{tabular}
\end{table}

\subsection{Robustness to Different Speech Rate}

\begin{figure}[!h]
    \vspace{-10pt}
    \centerline{\includegraphics[scale=0.85]{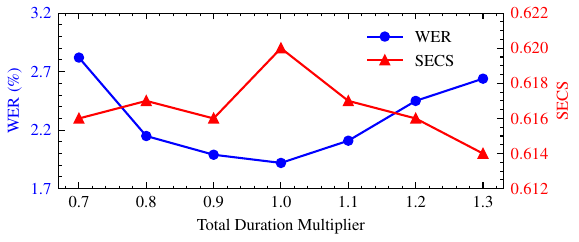}}
    \vspace{-10pt}
    \caption{
        WER and SECS of zero-shot text-to-speech with E1 TTS when scaling the predicted total duration by different factors.
    }
    \label{fig:dur_robust}
\end{figure}

To assess the robustness of E1 TTS to varying total durations, we measure the WER and SECS of zero-shot TTS results (as described in Section~\ref{sec:zstts}) while scaling the predicted duration by factors from 0.7 to 1.3. The results, shown in Figure \ref{fig:dur_robust}, demonstrate that E1 TTS can tolerate some error in total duration prediction.

\subsection{Robustness to Hard Cases}

We report the robustness of E1 TTS on challenging sentences containing difficult textual patterns for synthesis. We adopt the 100 hard sentences proposed in \cite{ELLAV} for evaluation. For each utterance, we randomly sampled 3-second-long speech prompts from LibriTTS test-clean. The results can be found in Table \ref{table:hard_cases}. We were surprised that E1~TTS$_\text{DMD}$ even outperforms StyleTTS 2 on this task, since StyleTTS 2 includes a duration predictor and therefore should not make alignment mistakes.

\begin{table}[!h]
\caption{Results of zero-shot text-to-speech WER on hard cases.}
    \label{table:hard_cases}
    \centering
    \begin{tabular}{lccccc}
        \toprule
        Model         & WER(\%)$\downarrow$ & Sub(\%)$\downarrow$ & Del(\%)$\downarrow$ & Ins(\%)$\downarrow$ \\
        \midrule 
        StyleTTS 2    & 4.83 & 2.17 & 2.03 & 0.61 \\
        CosyVoice     & 8.30 & 3.47 & 2.74 & 1.93 \\
        E1 TTS$_{\text{DMD}}$  & 4.29 & 1.89 & 1.62 & 0.74 \\
        \bottomrule
        \end{tabular}
\end{table}

\subsection{Sample Variation}

Adversarial training is known to be prone to mode collapse, leading to reduced sample diversity. The reverse KL divergence employed in DMD is also a mode-seeking loss. Therefore, it is important to investigate whether E1~TTS suffers from mode-dropping. We generated 100 random samples from zero-shot TTS models using the same target text~\cite{VITS}, ``How much variation is there?'', and an identical speech prompt\footnote{\texttt{LibriTTS, test-clean, 8555\_284449\_000044\_000000}}. We then computed the expected pairwise distances of MFCC and pitch \textit{sequences} between the generated samples. We applied dynamic time warping (DTW) to compute the distances. To measure duration variability, we estimated the phoneme durations using our RAD-TTS aligner and reported the expected pairwise Euclidean distance between the phoneme duration sequences. The results can be found in Table~\ref{table:diversity}. E1~TTS has higher sample diversity compared to StyleTTS 2 and CosyVoice, but it lags behind the autoregressive diffusion transformer-based system ARDiT$_\text{B=1, DMD}$.

\begin{table}[!h]
\caption{Expected pairwise sequence distance of samples.\\
(Higher values imply greater sample diversity.)}
    \label{table:diversity}
    \centering
    \begin{tabular}{lccc}
        \toprule
        Model      & MFCC$_{\times10^3}\uparrow$& Pitch$_{\times10^3}\uparrow$ & Duration$_{\times10^{-1}}\uparrow$\\
        \midrule 
        StyleTTS 2 & 6.2 & 9.2 & 2.8 \\
        CosyVoice & 7.3 & 7.9 & 4.2 \\
        ARDiT$_\text{B=1, DMD}$ & 11.7 & 10.6 & 6.0 \\
        E1 TTS$_\text{DMD}$ & 10.6 & 8.9 & 4.7 \\
        \bottomrule
        \end{tabular}
\end{table}

\section{Conclusions and Limitations}

In this study, we propose E1~TTS, an implicit-duration non-autoregressive (ID-NAR) TTS model with non-iterative~(1-step)~sampling capable of generating high-quality speech. Our experiments demonstrate that when trained on LibriTTS, E1~TTS can achieve comparable performance to strong NAR and AR baseline models. Despite relying only on implicit alignment and one-pass inference, E1~TTS generates diverse samples, and it is surprisingly robust to out-of-domain text inputs.

The inferior performance of E1 TTS$_\text{ODE}$ requires further investigation. SimpleSpeech~\cite{SimpleSpeech} observed poor sample quality when training a similar model on the LibriTTS dataset, indicating that the quantity of training data plays a crucial role. SESD~\cite{SimpleTTS2}~highlighted the significance of the noise schedule~\cite{NoiseScheduleBlog} in diffusion training and inference. The mode-seeking reverse KL divergence used in DMD may also contribute to the superior performance of E1~TTS$_\text{DMD}$, as it prioritizes sample correctness over distribution coverage, which aligns better with human perception.


\bibliographystyle{IEEEtran}
\bibliography{refs}

\end{document}